\documentclass[preprint,11pt, authoryear]{elsarticle}
\usepackage{amssymb}
\usepackage{graphicx}
\usepackage[margin=3.0cm]{geometry}

\usepackage{graphicx}
\usepackage{footnote}
\usepackage{threeparttable}

\begin{document}

\begin{frontmatter}

\title{Abundance analysis of the recurrent nova RS Ophiuchi (2006 outburst)}

\author{Ramkrishna Das \& Anindita Mondal}

\address{Astrophysics \& Cosmology Department, S N Bose National Centre for Basic Sciences, Salt Lake, Kolkata 700098, India\\
E-mail address: ramkrishna.das@bose.res.in, anindita12@bose.res.in}

\begin{abstract}
We present an analysis of elemental abundances of ejecta of the recurrent nova RS Oph using published optical and near-infrared spectra
during the 2006 outburst. We use the CLOUDY photoionization code to generate synthetic spectra by varying several parameters,
the model generated spectra are then matched with the observed emission line spectra obtained at two epochs. We obtain the best fit model parameters through the $\chi^{2}$ minimization technique. Our model results fit well with observed optical and near-infrared spectra. The best-fit model
parameters are compatible with a hot white dwarf source with T$_{BB}$ of 5.5 - 5.8 $\times$ 10$^{5}$ K and
roughly constant a luminosity of 6 - 8 $\times$ 10$^{36}$ ergs s$^{-1}$.
From the analysis we find the following abundances (by number) of elements with respect
to solar: He/H = 1.8 $\pm$ 0.1, N/H = 12.0 $\pm$ 1.0, O/H = 1.0 $\pm$ 0.4, Ne/H = 1.5 $\pm$ 0.1,
Si/H = 0.4 $\pm$ 0.1, Fe/H = 3.2 $\pm$ 0.2, Ar/H = 5.1 $\pm$ 0.1, and Al/H = 1.0 $\pm$ 0.1, all other elements were set at the solar abundance.
This shows the ejecta are significantly enhanced, relative to solar, in helium, nitrogen,
neon, iron and argon. Using the obtained parameter values, we estimate an ejected mass in the range
of 3.4 - 4.9 $\times$ 10$^{-6}$ M$_{\odot}$ which is consistent with observational results.
\end{abstract}


\end{frontmatter}
\section{Introduction}
RS Ophiuchi (RS Oph) is a well-observed recurrent nova (recurrence period $\sim$ 20 years) and is one of the
ten confirmed recurrent novae that belong to our galaxy (Kato $\&$ Hachisu 2012, Schaefer 2010).
The RS Oph system is composed of a massive ($\sim$ 1.35 M$_{\odot}$, Kato, Hachisu $\&$ Luna 2008) white dwarf (WD) primary accompanied by a red giant secondary of estimated spectral class around M2 III (Worters et al. 2007 and references therein). Brandi et al. (2009) estimated the orbital period to be 453.6 days, the red giant mass, M$_{g}$ = 0.68 - 0.80 M$_{\odot}$ and the orbital inclination, i = 49$^{\circ}$ - 52$^{\circ}$ for the system.
The outburst takes place due to a thermonuclear runaway (TNR) on the WD surface
that accretes matter from the secondary red giant companion. The outburst causes ejection of mass $\sim$ 10$^{-6}$ - 10$^{-8}$ M$_{\odot}$
at a high speed of $\sim$ 4000 km s$^{-1}$ (e.g. Buil 2006). Previous studies of outbursts indicate that
the WD mass of RS Oph is possibly increasing due to the accumulation of a percentage of the accreted matter on its surface.
Consequently, the mass of the WD in RS Oph may gradually reach the Chandrasekhar limit and explode as a Type Ia supernova -
this has made RS Oph an object of great significance to the astrophysicists. However, there have been considerable debates about
this hypothesis (Starrfield et al. 2004, Wood-Vasey $\&$ Sokoloski 2006). \\

RS Oph was detected in outburst previously in 1898, 1933, 1958, 1967, 1985; its latest outburst was discovered on 2006
February 12.83 UT (Hirosawa 2006). The reason for the much shorter recurrence period in the RS Oph system,
in comparison to classical novae (CNe),
is due to combined effect of the high WD mass and a high accretion rate (Starrfield et al. 1985, Yaron et al. 2005).
The two recent outbursts of RS Oph in 1985 and 2006 have been observed intensively over a wide range of wavelengths,
from X-rays to the radio regions (Bode 1987;  Evans, Bode, O'Brien $\&$ Darnley 2008). Detailed studies
display very similar characteristics of the outbursts. In the early phase of the outburst, the spectra show broad, low-ionization emission features
of H, He, N, O and Fe; the nova enters quickly (about a month after outburst) to the nebular phase with the
emergence of strong coronal (e. g., [Fe XIV] 0.5303 $\mu$m,  [Ar X] 0.5535 $\mu$m, [Fe X] 0.6374 $\mu$m, [Si VI] 1.9641 $\mu$m,
[Al IX] 2.0444 $\mu$m, [Mn XIV] 2.0894 $\mu$m) and nebular lines (e.g., [O III] (0.4363, 0.4959 and 0.5007 $\mu$m) and
[N II] 0.5755 $\mu$m) (Iijima 2009, Banerjee et al. 2009). The nova light curves also behave similarly; they decline fast with $t_{2}$ $\sim$ 6
and $t_{3}$ $\sim$ 17 days (Rosino 1987, Munari et al. 2007).  The key result of the 1985 and 2006 observations was the detection of
a shock that is generated while the ejecta interacts with the surrounding wind of the red giant secondary (Bode $\&$ Kahn
1985, Das et al. 2006 and references therein) and a non-spherical bi-polar shape of the nova ejecta (e.g., Taylor
et al. 1989, Chesneau et al. 2007, Bode et al. 2007).
Further investigations also helped to determine a few important parameters viz. determinations of the distance,
d = 1.6 $\pm$ 0.3 kpc (Hjellming et al. 1986), the interstellar hydrogen column density, N $\sim$ 2.4 $\times$ 10$^{21}$ cm$^{-2}$ (Hjellming et al. 1986), and an interstellar reddening of E(B - V) = 0.73 (Snijders 1987).\\

However, despite plenty of observations of RS Oph, the abundance analysis of the nova ejecta has not been done adequately. A Few values
have been calculated, for example, from optical studies of the 1985 outburst, Anupama $\&$ Prabhu (1985)
derived a helium abundance of $n(He)/n(H)$ = 0.16; Evans et al. (2007) estimated the O/Ne ratio
(by number) to be  $\gtrsim$ 0.6 from IR
studies of the 2006 outburst. A complete knowledge of elemental abundances in the
ejecta is of crucial importance for several reasons, for example, to understand the TNR process that leads to the nova explosion,
the composition of material of the WD\textbf{,} as there is a possibility of mixing of WD material with the ejecta,
the contribution of novae to the chemical evolution of galaxy etc. In this paper, we report the results of an elemental abundance
analysis of the ejecta of RS Oph by modeling its available optical and near-infrared (NIR) spectra observed during the 2006
eruption. We have used the photoionization code CLOUDY (version 13.02; Ferland et al., 2013) to generate spectra, by varying
the parameter values. Model generated spectra are then compared with the observed emission line spectra, the best fit model is
chosen by calculating the corresponding $\chi ^{2}$ values. The procedure of modeling is described in section 3; results obtained from the analysis
is described in section 4.

\section{Photoionization model analysis}
We use the CLOUDY photoionization code,  C13.02 (Ferland et al., 2013) for the abundance analysis in RS Oph.
The benefit of using photoioinization models is that in addition to elemental abundances, they also provide
estimate of several other parameters, e.g., density, source luminosity, source temperature etc. Previously, this method
was used to determine the elemental analysis and physical characteristics of a few novae by modeling the observed spectra,
for example, LMC 1991 (Schwarz 2001), QU Vul (Schwarz et al. 2002), V1974 Cyg (Vanlandingham et al. 2005),
V838 Her $\&$ V4160 Sgr (Schwarz et al. 2007a), V1186 Sco (Schwraz et al. 2007b), V1065 Cen (Helton et al. 2010).
The photoionization code CLOUDY uses a set of parameters that specify the initial physical conditions of the source and the ejected shell.
The source is described by the spectral energy distribution of the continuum source, its temperature and luminosity. The physical condition
of the shell is described by the density, inner and outer radii, geometry, covering factor (fraction of 4$\pi$ sr enclosed by the model shell),
filling factor (ratio of the contribution of the dense shell to the diffuse shell) and elemental abundances (relative to solar).
The density of the shell is set by a hydrogen density parameter and the elemental abundances, relative to hydrogen, are set by the abundance
parameters. The hydrogen density, n(r), and filling factor, f(r), may vary with the radius as given by the following relations,

\begin{equation}
n(r) = n(r_{0}) (r/r_{0})^{\alpha} cm^{-3} \,\,\,\,\, \,\,\&\,\,\,\,\,\, f(r) = f(r_{0}) (r/r_{0})^{\beta}
\end{equation}

where, r$_{o}$ is the inner radius, $\alpha$ and $\beta$ are exponents of power laws.
We choose $\alpha$ = -3, the filling factor = 0.1  and the filling factor power-law exponent, ($\beta$) = 0, which
are the typical values used in similar kind of studies (e.g., Schwarz 2002, Vanlandingham et al. 2005, Helton et al. 2010).\\

CLOUDY solves the equations of thermal and statistical equilibrium using the above mentioned set of input parameters to
generate output spectra from the non-LTE ejecta illuminated by the central source. Its calculations incorporate effects of
important ionization processes, e.g., photo, Auger, collisional $\&$ charge transfer and recombination
process viz. radiative, dielectronic, three-body recombination, and
charge transfer. We assume the continuum shape to be a blackbody of a high temperature $T_{BB}$ $\ge$ 10$^{5}$ K,
as done in the previous investigations, to ensure that it supplies the correct amount of photons for photoionization.
The output predicts the flux of emission lines, which is compared to the measured line
fluxes in the observed spectra.\\

\begin{table*}
\small
\centering
\caption{A log of the used spectroscopic data of RS Oph (2006).}
\smallskip
\begin{threeparttable}
\centering
\begin{tabular}{l c c c c c c c}
\hline\
Epochs & Wavelength    &     Date of           &    Days after       &		Telescopes / 		  &	Resolution & Refer-  \\
       &    band       &     observation       &     outburst        &		 Instruments		  &	           & ences   \\
\hline
D31    & Optical       &  March 15           &     31    &  Observatorio Astr\'{o}nomico          & $\sim$ 1000 -  &	1    \\
       &               &                     &           &  Nacional en San Pedro      & 3500           &         \\
       &               &                     &           &  M\'{a}rtir/ Boller \& Chivens      &                      \\
       &  NIR          &  March 16         &     32  &  Mt. Abu Telescope/ NICMOS                 & $\sim$ 1000    &    2    \\
\hline
D49    &  Optical      &  April 4          &     50    & Astrophysical Observatory of 		      & $\sim$ 1000    &	3    \\
       &               &                    &           &  Asiago/ Boller \& Chivens              &                          \\
       &  NIR          &  April 2   &     49  &      Mt. Abu Telescope/ NICMOS                    & $\sim$ 1000    &    2    \\
\hline
\end{tabular}
\begin{tablenotes}
\item[] 1 = Riberio et al. 2009; 2 = Das et al. 2006 $\&$ Banerjee et al. 2009; 3 = Iijima 2009
\end{tablenotes}
\end{threeparttable}
\end{table*}

\section{Modeling procedure}
For the present analysis, we use observed optical and NIR $JHK$ spectra of the 2006 outburst of RS Oph. Moedeling of both optical and NIR
data enables to sample over a broader range of ionization and excitation levels in the emission lines and thus helps to constrain
the results more accurately. We choose two epochs of observations taken at different times of the nova evolution, that had nearly simultaneous
optical and NIR spectra and form two data sets represented by D31 and D49.
D31 consists of optical and NIR spectra that is observed, respectively, on 2006 March 15 $\&$ March 16 i.e. approximately 31 days after outburst;
whereas, D49 consists of optical and NIR spectra observed, respectively, on 2006 April 4 $\&$ April 2, i.e. approximately 49 days after outburst.
Here, for simplicity, we assume that the physical condition and corresponding parameters, in the ejecta remain unchanged over 1-2 days.
Details about the used spectra are presented in Table 1.
A detailed modeling using more data sets extended over a larger time period and including other wavelength regions
is in progress and will be published later.\\

We assume a spherically symmetric expanding shell geometry of the ejecta that is illuminated by the central source.
Several spectra are generated by varying the free parameters, one after one, viz. hydrogen density, underlying luminosity,
effective blackbody temperature and abundances of only those elements
which showed observed lines. The abundances of other elements, which do not show any emission line, were
fixed at solar values. Since, novae ejecta are not homogenous in density, we assume that the ejecta is composed of at least two different density
regions - one for the higher density to fit the lower ionization lines and the other for the lower density to fit the higher ionization lines.
To reduce the number of free parameters in the final model, each component is subjected to the same parameters except
the hydrogen densities at the inner radius and the covering factors assuming that the sum of the two covering factors be less than or equal to 1. Thus,
the overall number of free parameters increases by 2 due to the second component's initial density and covering factor.
The final model line ratios were calculated by adding the line ratios of each component after multiplying by its covering factor.
Thus, this method is only a first-order approximation to incorporate density gradients into the photoionization
analysis; there is no coupling or interaction between the components, as expected in reality.
However, this procedure has been used successfully in the case of other novae, e. g., V838 Her (Schwarz et al. 2002), V1974 Cyg (Vanlandingham et al. 2005), V838 Her $\&$ V4160 Sgr (Schwarz et al. 2007a), V1186 Sco (Schwraz et al. 2007b)  and V1065 Cen (Helton 2010).\\

Since CLOUDY uses a large number of parameters and many of the parameters are interdependent, it is
difficult to determine the uniqueness of any solution by checking the generated spectra visually. Hence, we choose the the best fit model
by calculating $\chi^{2}$ and reduced $\chi^{2}$ ($\chi^{2}_{red}$) of the model given by,

\begin{equation}
\chi ^{2} = \sum\limits_{i=1}^n (M_{i} - O_{i})^{2}/ \sigma_{i}^{2}, \,\,\,\,\&\,\,\,\, \chi^{2}_{red} = \chi^{2}/\nu
\end{equation}

where, $n$ = number of observed lines, $n_{p}$ = number of free parameters, $\nu$ = degree of
freedom = n - n$_{p}$, M$_{i}$ = the modeled ratio of line flux to hydrogen line flux, O$_{i}$ = measured flux ratio,
and $\sigma_{i}$ = error in the observed
flux ratio. We estimate an error in the range of 10 - 30$\%$, depending upon the strength of a
spectral line relative to the continuum, possibility of blending with other lines,
and formal error in the measurement of line flux. A model is considered good if the value of $\chi^{2}$ $\sim$ $\nu$ such that the
($\chi^{2}_{red}$) value is low (typically in the range of 1 - 2).\\

To minimize the number of free parameters, the inner ($R_{in}$) and outer ($R_{out}$) radii of the ejected shell are held constant during the
iterative process of fitting the spectra. For simplicity,
we assume a spherical geometry of the expanding shell with ($R_{in}$) and  ($R_{out}$) defined by the minimum ($V_{min}$) and maximum ($V_{max}$) expansion velocities, respectively. We adopt $V_{min}$ = 3500 km s$^{-1}$ and $V_{max}$ = 4500 km s$^{-1}$ during explosion based on values calculated from the optical and NIR emission lines (e.g., Das et al. 2006, Skopal et al. 2008). The velocities remain constant during the free expansion phase that lasted for first $\sim$ 3 days (taking an average of 2 and 4 days, derived by Sokoloski et al. 2006 and Das et al. 2006, respectively) and then decrease gradually with time (t) as $t^{-0.6}$ (Das et al. 2006) due to interaction with surrounding matter from the red giant companion. Following this process, we calculate and find $V_{min}$ $\sim$ 1950 km s$^{-1}$ and $V_{max}$ $\sim$ 2500 km s$^{-1}$ on day 5.5. These values are consistent with the interferometric results (Chesneau et al., 2007)  that show the evidence of two different radial velocity fields inside the ejecta, with $V_{min}$ $\le$ 1800 km s$^{-1}$ and $V_{max}$ $\sim$ 2500 - 3000 km s$^{-1}$, at day 5.5. The agreement between the two results gives us confidence about the model we adopted here. Proceeding further, we calculate the velocities and expansion of the shells on each date, then add them up to calculate the final values of $R_{in}$ and $R_{out}$ on D31 and D49.\\

\begin{table*}
\tiny
\centering
\caption{Observed and Best-Fit Cloudy Model Line Fluxes $^{a}$.}
\smallskip
\begin{threeparttable}
\centering
\begin{tabular}{l c c c c c c c}
\hline
       &				        &	D31	   &			 &	   	      & 		  &  D49    &           \\
Line       & $\lambda$ ($\mu$ m)& Observed &  Modeled    & $\chi^{2}$ &	 Observed & Modeled & $\chi^{2}$\\
\hline
           &                    &          & Optical    &              &            &         &           \\
\hline
$[$Fe VII] &	0.3760			&	... 	&	 ...	&	...		    &	0.21	&	0.12	&	0.80 \\
$[$Ne III]  & 0.3868          	&   ...  	&  	...		&   ...   		&  	0.32    &  	0.32 	&   0.00   \\
H I, He I	& 0.3889            &   ... 	&    ...	&  ...   		&   0.43    &  	0.36	&   0.49   \\
H $\epsilon$, [N III] & 0.3970 &	...		&	...		&	...		    &	0.24	&	0.45	&	4.40 \\
He I, He II	& 0.4029 			&	0.01	&	0.05	&	0.12		&	0.16	& 	0.05	& 	1.21\\
H $\delta$& 0.4103				&	0.14	&	0.20	&	0.34		&   0.51 	&	0.31	& 	3.99 \\
Fe II, [Ni XII]    &	0.4233	&	0.02	&	0.17	&	2.02		& 	0.14 	& 	0.07 	& 	0.42 \\
H $\gamma$& 0.4341 				&	0.38	&	0.40	&	0.04		&	0.50 	& 	0.55 	& 	0.25\\
$[$Fe II] 	& 0.4415 			&	0.02	&	0.10	&	0.65		&	0.07 	& 	0.07 	& 	0.00 \\
He I 	& 0.4472				&	0.05	&	0.12	&	0.56		&	0.11 	& 	0.10 	& 	0.01 \\
Fe II	& 0.4523 				&	0.04	&	0.10	&	0.46		&	0.08	&	0.06 	& 	0.04\\
Fe II 	& 0.4549 				&	0.04	&	0.11	&	0.47		&	0.08 	& 	0.06 	& 	0.04\\
Fe II	& 0.4590 				&	0.04	&	0.11	&	0.45		&	0.08 	& 	0.04 	& 	0.16\\
He II	& 0.4666 				&	0.35	&	0.57	&	4.60		& 	0.73	& 	0.79 	& 	0.36 \\
H $\beta$	& 0.4863 			&	1.00   	&	1.00 	&	0.00		& 	1.00	& 	1.00	& 	0.00 \\
He II, Fe II	& 0.4923 		& 	0.07	&	0.11	&	0.18  	    &	0.15	&	0.07	&	0.64 \\
$[$O III], He I, Fe II& 0.5017 	& 	0.23	&	0.46	& 	5.29		& 	0.43 	& 	0.29 	& 	1.96 \\
$[$Fe II], [Fe VII] &	0.5158	&	0.08	&	0.18	&	0.92		&	0.16	&	0.28	&	1.44 \\
Fe II & 0.5235 					&	0.03	&	0.04	&	0.01		&	...		&	...		&	... \\
Fe II & 0.5276					&	0.04	&	0.15	&	1.12		&	...		&	...		&	... \\
Fe II       & 0.5317			&	0.01	&	0.05	&	0.14		&	...		&	...		&	...	\\
He II	    & 0.5411 			&	0.01	&	0.10  	& 	0.25		&	0.08	& 	0.10 	& 	0.03\\
$[$Ar X]	& 0.5535 		    & 	0.05	& 	0.08 	& 	0.11		& 	0.14 	& 	0.6		& 	0.64\\
$[$N II]	& 0.5755 		    & 	...	    & 	...    	& 	... 	    & 	0.12 	& 	0.16	& 	0.16\\
He I		& 0.5876 			& 	...		& 	...		& 	...		    & 	0.42 	&   0.43	& 	0.01\\
$[$Fe X]	& 0.6374 			& 	...		& 	...		& 	...		    & 	0.33 	& 	0.14	& 	3.61\\
He I		& 0.6678 			& 	...		& 	...		& 	...		    & 	0.23	& 	0.13 	&  	1.00\\
$[$Ar XI]   & 0.6919 			& 	...		& 	...		& 	...		    & 	0.10	& 	0.04 	& 	0.36\\
He I        & 0.7065 			& 	...		& 	...		& 	...		    & 	0.22 	& 	0.40	& 	3.20\\
He I        & 0.7281			& 	...		& 	...		& 	...		    & 	0.13 	& 	0.09    & 	0.16\\
\hline
&&& J-band & & & & \\
\hline
He I		& 	1.0830		    &	4.39		 & 	 4.70	& 	 2.40	& 	11.80 	&  11.45 	& 	3.06 \\
Pa $\gamma$	& 	1.0938			& 	1.02		 & 	 0.78	& 	 1.44	& 	1.79 	& 	0.85  	& 	9.82 \\
He II   	&  	1.1626 			& 	0.11		 & 	 0.10	& 	 0.02	& 	0.40  	& 	0.26 	& 	1.96 \\
Pa $\beta$	& 	1.2818			&	1.00		 &	 1.00 	& 	 0.00	&	1.00 	& 	1.00 	& 	0.00 \\
\hline
&&& H-band & & & & \\
\hline
Br 20	& 1.5184 				&	0.57		&	0.50		&	0.49		&	0.33		&	0.29 	    &	0.16\\
Br 19	& 1.5256 				&	0.31		&	0.24		&	0.49		&	0.38		&	0.25		&	1.69\\
Br 18	& 1.5341 				&	0.45		&	0.37		&	0.64		&	0.47		&	0.38		&	0.81\\
Br 17	& 1.5439 				&	0.45		&	0.31		&	1.96		&	0.60		&	0.38		&	4.84\\
Br 16	& 1.5570 				&	0.71		&	0.63		&	0.64		&	0.91		&	0.73		& 	3.24\\
Br 15	& 1.5685 				&	0.66		&	0.53		&	1.69		&	0.60		&	0.60		& 	0.00\\
Br 14	& 1.5881 				&	0.76		&	0.69		&	0.49		&	0.77		&	0.70		& 	0.49\\
Br 13	& 1.6109				&	0.82		&	0.55		&	1.82 	&	0.69		    &	0.59		& 	1.00\\
Br 12	& 1.6407 				&	1.00		&	1.00		& 	0.00	& 	1.00		    &	1.00 		& 	0.00\\
Br 11	& 1.6806 				&	1.81		&	1.82 	    &	0.01 	&	1.34 	        &	1.20		& 	1.96\\
Fe II	& 1.6872 				&	1.01 	    &	0.91		&	0.99 	&	0.91 	        &	0.78		& 	1.69\\
He I	& 1.7002 				&	0.69 	    &	0.79		&	1.00	&	0.74 	        &	0.90  	    & 	2.56\\
Br 10	& 1.7362 				&	2.18		&	2.26		&	0.64 	&	2.10  	        &	2.26		& 	2.56\\
Fe II   & 1.7406 				&  	0.34		&	0.28		&	0.36 	&	0.69		    &	0.58		&	1.21\\
\hline	
&&& K-band & & & & \\
\hline
Br 8 	& 1.9446 				&	 ...	&    ...	 	& ...	 	&	0.8		    & 	0.86		& 	0.36 \\
$[$Si VI]	& 1.9621 			&	0.46	&	0.29		&	2.28 	&	1.3		    &	1.80		&	0.64 \\
$[$Al IX]   & 2.0444 		    &	0.04	&	0.08		&	0.13 	&	0.13		&	0.10		& 	0.09 \\
He I		& 2.0581 			&	0.71	&	0.44		&	1.80	&	0.87		&	0.40		& 	5.50 \\
He I 	    & 2.1120 			&	0.14	&	0.08 	    &	0.36 	&	0.13	    &	0.12		& 	0.04 \\
Br $\gamma$&	2.1655 			&	1.00	&	1.00 		& 	0.00	&	1.00		&	1.00 		& 	0.00 \\
\hline
Total &							&			& 			    & 	37.99 	& 			    & 			    &	69.06 \\
\hline
\end{tabular}
\begin{tablenotes}
\item[a] Relative to H $\beta$ in optical, Pa $\beta$ in $J$ band, Br 12 in $H$ band and Br $\gamma$ in $K$ band.
\end{tablenotes}
\end{threeparttable}
\end{table*}

\begin{figure}
\centering
\includegraphics[bb= 265 80 1280 460, width=6.0 in, height = 2.6 in, clip]{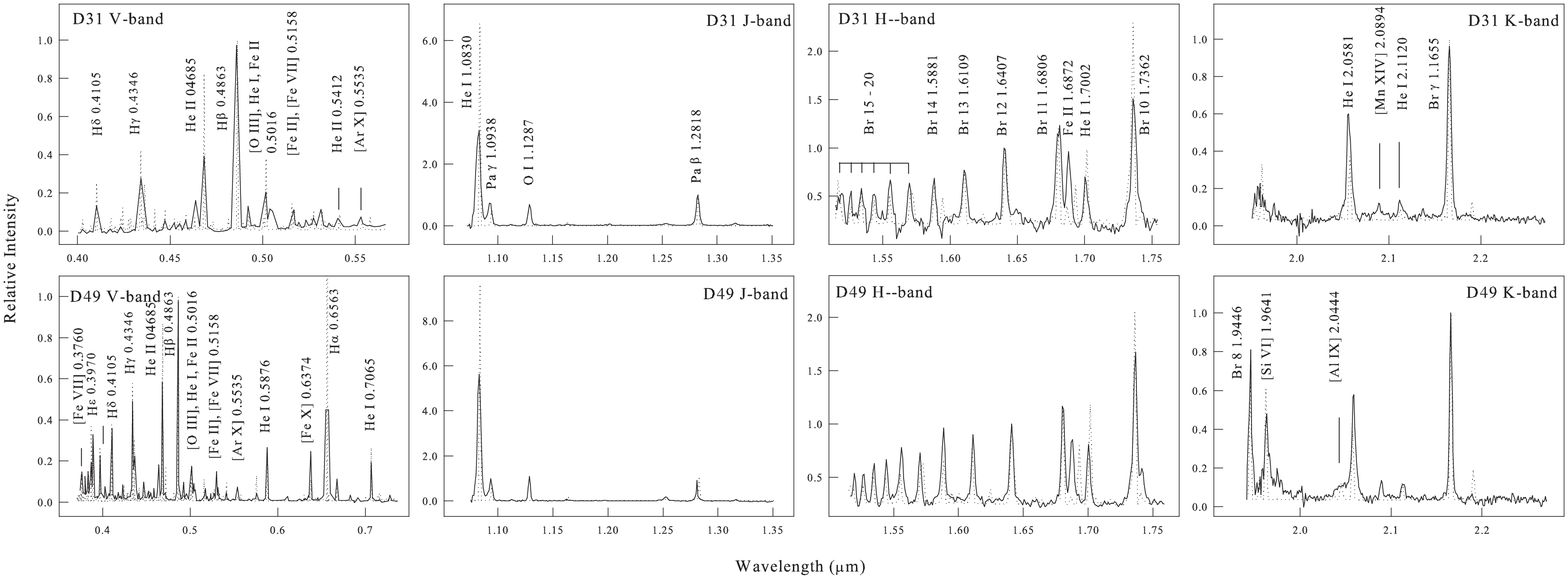}
\caption{Best Cloudy model (dotted line) fit to the observed optical and NIR $JHK$ spectra (solid lines) of RS Oph observed during the epochs D31 and D49 during the 2006 outburst. The spectra were normalized to H $\beta$ in optical, Pa $\beta$ in $J$ band, Br 12 in $H$ band and Br $\gamma$ in $K$ band. Also few of the strong features have been marked (see text for details).}
\end{figure}


\section{Results}
The results of our analysis are presented in Table 2 $\&$ 3. Table 2 shows the relative fluxes of the best-fit model predicted and observed lines
along with corresponding $\chi^{2}$ values. Here, we have
considered only the lines which are present both in the model\textbf{-}generated and observed spectra. We assume a
distance of 1.6 kpc for RS oph to match the predicted luminosities with the
reddening-corrected observed flux. The measured line fluxes were dereddened using E(B - V ) = 0.73 (Snijders 1987)
and compared to the output of each CLOUDY model to calculate $\chi ^{2}$ for the fit.
The line intensities are determined from direct integration of the line; for features with multiple components, the profiles were decomposed
with multiple Gaussians using IRAF tasks. To minimize the errors associated with flux calibration
between different epochs and wavelength regions, we calculate the modeled and observed flux ratios relative to prominent hydrogen lines, which are not blended with any other features, within a given wavelength region, e. g., relative to H$\beta$ in the optical region, relative to Paschen $\beta$ in the $J$-band, relative to Brackett 12 in the $H$-band and relative to Brackett $\gamma$ in the  $K$-band.\\

\begin{table*}
\centering
\caption{Best-fit CLOUDY Model parameters}
\smallskip
\begin{threeparttable}
\centering
\begin{tabular}{l c c c c c c c}
\hline
 Parameters                                             &		D31		  &	D49	 & Predicted Abundances    \\
\hline\\
$T_{BB}$ ($\times$10$^{5}$ K)                           &     5.8         &   5.5    & ...      \\
Source luminosity ($\times$10$^{36}$ erg s$^{-1}$)		&    6.3          &   8.0    & ... \\
Clump hydrogen density ($\times$10$^{8}$ cm$^{-3})$     &   10.0          &   6.3    & ...\\
Diffuse hydrogen density ($\times$10$^{8}$ cm$^{-3})$   &    1.6          &   1.0    & ...\\
$\alpha$\tnote{a}	                                    &   	-3	      &	-3	     & ...\\
Inner radius ($\times$10$^{14}$ cm )                    & 	2.1           &	2.8      & ...\\
Outer radius ($\times$10$^{14}$ cm)                     &    4.8          & 6.8      & ...\\
Clump to diffuse covering factor                        &    90/10        & 85/15       & ... \\
Filling factor	                                        &	0.1	          &	  0.1	    & ... \\
$\beta$\tnote{b}		                                &	0.0	          &	  0.0	        & ... \\
He/He$_\odot$\tnote{c}                                  &   1.8 (11)      &   1.9 (16)    & 1.8 $\pm$ 0.1\\
N/N$_\odot$                                             &    ...          &   12.0 (2)    & 12.0 $\pm$ 1.0 \\
O/O$_\odot$                                             &   1.0 (1)       &   1.0 (1)     & 1.0 $\pm$ 0.4 \\
Ne/Ne$_\odot$		                                    &	...   	      &	1.5	(1)	     & 1.5 $\pm$ 0.1 \\
Si/Si$_\odot$                                           &   0.3 (1)       &   0.5 (1)    & 0.4 $\pm$ 0.1\\
Fe/Fe$_\odot$                                           &   3.0 (12)      &   3.5 (11)   & 3.2 $\pm$ 0.2 \\
Ar/Ar$_\odot$                                           &   4.9 (1)       &   5.2 (2)    & 5.1 $\pm$ 0.1\\
Al/Al$_\odot$                                           &   0.9 (1)       &   1.1 (1)   & 1.0 $\pm$ 0.1\\
Ejected Mass ($\times$10$^{-6}$ M$_\odot$)              &   3.4           & 	4.9     & ...\\
Number of observed lines (n)                            &    42           &   51       & ... \\
Number of free parameters (n$_{p}$)                     &    11           &   13       & ... \\
Degrees of freedom ($\nu$)		                        &	 31	          &	  38	   & ... \\
Total $\chi^{2}$ 	                                    &	38.0          &	  69.1     & ... \\
$\chi^{2}_{red}$                                        & 	1.2	          &	   1.8	   & ... \\
\\
\hline
\end{tabular}
\begin{tablenotes}
\item[a] Radial dependence of the density $r^{\alpha}$
\item[b] Radial dependence of filling factor $r^{\beta}$
\item[c] Abundances are given in logarithmic scale, relative to hydrogen. All other elements which are not listed
in the table were set to their solar values. The number in the parentheses represents number of lines used in
determining each abundance.
\end{tablenotes}
\end{threeparttable}
\end{table*}

The best fit modeled spectra (dotted line) together with  the observed optical and NIR
spectra (solid lines) on D31 and D49, are shown in Figure 1.  We have marked the strong spectral lines on the figure, a
detailed observed line list is presented in Banerjee et al. (2009) and (Iijima 2009).
The spectra on D31 are dominated by prominent features of low ionization lines of H$\epsilon$ (0.3970 $\mu$m), H$\delta$ (0.4103 $\mu$m), H$\gamma$ (0.4341 $\mu$m), He II (0.4666 $\mu$m), H $\beta$ (0.4863 $\mu$m) in the optical region;  He I (1.0830 $\mu$m), Pa $\gamma$ (1.0938 $\mu$m), O I (1.1287 $\mu$m), Pa $\beta$ (1.2818 $\mu$m), Brackett series lines, Fe II (1.6872 $\mu$m), He I (1.7002 $\mu$m) in the $H$-band; and
He I (2.0581 $\mu$m), Br $\gamma$ (2.1655 $\mu$m) in the $K$-band. There are also higher ionization lines, viz. [Fe VII] (0.3760 $\mu$m), a blended feature of [O III], He I $\&$ Fe II (0.5017 $\mu$m), \textbf{a} blended feature of [Fe II] $\&$ [Fe VII] (0.5158 $\mu$m), [Ar X] (0.5535 $\mu$m), [Fe X] (0.6374 $\mu$m)
in the optical; and [Si VI] (1.9641 $\mu$m), [Al IX] (2.0444 $\mu$m), [Mn XIV] (2.0894 $\mu$m) in the NIR region.
As mentioned in the earlier section, a single shell of ejecta could not generate all of these lines.
For example, if we consider a clumpy shell of high density, e.g. 10$^{9}$ cm$^{-3}$ only, the modeled spectra fit the majority of the
lines in the observed spectra, but systematically under represents the higher ionization lines, for example, [Al IX] (2.0444 $\mu$m), Fe II (1.7406$\mu$m), [Fe X] (0.6374 $\mu$m) and [Fe VII] (0.3760 $\mu$m). This one component shell also increases the strength of the K-band He I line enormously. On the other hand, only a diffuse shell of lower density can produce the higher-ionization lines successfully but can not produce the He I and Fe II lines sufficiently to match the observed lines. Hence, for a better fitting of the lines, it was required to consider a two-component model consisting of a dense shell (hydrogen densities, by number: 10 x 10$^{8}$ cm$^{-3}$ and 6.3 x 10$^{8}$ cm$^{-3}$ on D31 and D49, respectively) and a diffuse shell
(1.6 x 10$^{8}$ cm$^{-3}$ and $\&$ 1.0 x 10$^{8}$ cm$^{-3}$ on D31 and D49, respectively). The ratio of the
clump to diffuse components are 90:10 and 85:15, respectively, on D31 and D49. This
indicates that the ejecta volume was dominated more by the dense gas.
The best-fit models use a blackbody temperature (T$_{BB}$) of  5.8 $\times$ 10$^{5}$ K and 5.5 $\times$ 10$^{5}$ K and
a luminosity of 6.3 $\times$ 10$^{38}$ ergs s$^{-1}$ and 8.0 $\times$ 10$^{38}$ ergs s$^{-1}$, respectively, on D31 and D49.
This is in agreement with the WD temperature of around 8 $\times$ 10$^{5}$ K derived from x-ray studies by Nelson et al. (2008).
The low $\chi^{2}_{red}$ values of 1.2 and 1.8 for the two dates, respectively, indicate that the fits are satisfactory.
The best-fit model parameters for each epoch are described in Table 3.\\

Despite the low values of $\chi^{2} _{red}$, the model still has some problems.
An inspection of Table 2 shows that a few lines make the highest contributions to the total $\chi^{2}$, for
example, He II (0.4666 $\mu$m), the blended feature of [O III], He I $\&$ Fe II (0.5017 $\mu$m)
on D31 and H $\epsilon$ $\&$ [N III] (0.3970 $\mu$m), H $\delta$, [Fe X] (0.6374 $\mu$m),
He I (0.7065, 1.0830 and 2.0581 $\mu$m), Pa $\gamma$, Br 16 and 17 on D49. The possible reason for these misfits
is that a two-component model is not sufficient to adequately reproduce the complex density structure of the nova ejecta;
consequently, the exact conditions inside the ejecta in which these lines are formed, have not been not reproduced correctly.
In addition, CLOUDY was unable to adequately reproduce a few of the observed lines, e.g., Si II (0.5041, 0.5056 $\mu$m),
Fe II (0.5235, 0.5276, 0.5317 $\mu$m), [Fe XIV] (0.5303 $\mu$m), N III (0.4641 $\mu$m), Raman emission band (0.6830 $\mu$m),
He I (1.1969, 1.2527 $\mu$m), N I (1.2074, 1.2096, 1.2470 $\mu$m), O I (1.1287, 1.3164 $\mu$m), [Mn XIV] (2.0894 $\mu$m) and He I (2.1120 $\mu$m).
So, we have excluded these lines from our analysis. The problem of reproducing the NIR O I emission
lines (1.1287 $\mu$m and 1.3164 $\mu$m) and He I (2.0581 $\mu$m) has also been reported by
Helton et al. (2010). They attributed these difficulties to related excitation mechanisms which are not properly
included in the present CLOUDY code.\\

Overall abundance values for RS Oph, calculated from the mean of these modeled results, are presented in Table 3. Abundances are
given as logarithm of the numbers relative to hydrogen and relative to solar.
The derived values show that the helium, nitrogen, neon, iron and argon abundances are all enhanced relative to solar, while
the oxygen and aluminium abundances are solar with respect to hydrogen, and silicon abundances are subsolar with respect to hydrogen.
We note that there is no prominent line of oxygen in the observed spectra. Our determination of the oxygen abundance is based on the feature
at 0.5017 $\mu$m that is blended with He I and Fe II. Therefore the calculated abundance value of oxygen may incorporate a significant error. Similarly, the calculation of abundances of silicon, argon and aluminium is based on one or two observed lines only. Modeling of multi-wavelength
spectra, observed over a longer time scale may improve the level of accuracy.

\subsection{Estimation of \textbf{the} ejecta mass}
Using the parameter values determined from the present analysis, we can estimate the hydrogen ejected mass predicted by the two component
models using the following relation (e.g., Schwarz 2001, 2002),

\begin{equation}
M_{shell} = n(r_{0}) f(r_{0}) \int \limits_{R_{in}}^{R_{out}} (r/r_{0})^{\alpha + \beta} 4\pi r^{2} dr
\end{equation}

The resulting mass is then multiplied by corresponding covering factors to obtain mass of the dense and diffuse shells;
the final mass is calculated by adding the masses of these two shells. Using this method, we find an ejected mass
of 3.4 $\times$ 10$^{-6}$ M$_{\odot}$ and 4.9 $\times$ 10$^{-6}$ M$_{\odot}$
for D31 and D49, respectively. Similar values of 1 - 3 $\times$ 10$^{-6}$ M$_{\odot}$ and 3 $\times$ 10$^{-6}$ M$_{\odot}$ were estimated
from other studies (e.g., Das et al. 2006; Kato, Hachisu $\&$ Luna 2008) of the 2006 outburst of RS Oph.
From the light curve analysis, Hachisu $\&$ Kato (2001) estimated an ejecta mass of $\sim$ 2 $\times$ 10$^{-6}$ M$_{\odot}$ for a WD of mass
of 1.35 M$_{\odot}$. Therefore the present result also favors the possibility of a high mass of $\sim$ 1.35 M$_{\odot}$ in the RS Oph system.

\section{Summary}
We have used the photoionization code CLOUDY to model the observed optical and NIR emission-line spectra of RS Oph observed on two epochs, 31 and 49 days after the outburst, during 2006. We generated a set of spectra by varying several parameters and assuming a spherical geometry of the ejecta that consists of two different shells of different densities. The model\textbf{-}generated spectra were then compared with observed spectra, the best fit parameters were chosen using the $\chi^{2}$ technique. The best-fit model parameters are in agreement with a hot WD with temperature of $\sim$ 5.5 - 5.8 $\times$  10$^{5}$ K and luminosity of 6 - 8 $\times$ 10$^{36}$ ergs s$^{-1}$. The abundance analysis shows that the ejecta are significantly enhanced, relative to solar, in helium, nitrogen, neon, iron and argon as well as silicon was found to be subsolar. We estimate an ejected mass in the range of 3.4 - 4.9 $\times$ 10$^{-6}$ M$_{\odot}$ which fits a high mass $\sim$ 1.35 $M_{\odot}$ of the associated WD in the RS Oph system. In continuation of the present work, a complete and detailed analysis of abundances and related parameters, covering a larger wavelength region and evolution period, will be the subject of a future work.

\section*{Acknowledgments}
The research work at S N Bose National Centre for Basic Sciences is funded by the Department
of Science and Technology, Government of India.

\end{document}